\documentstyle[12pt,twoside,fleqn,espcrc1,psfig]{article}

\newcommand{\be}{\begin{equation}}
\newcommand{\ee}{\end{equation}}
\newcommand{\bea}{\begin{eqnarray}}
\newcommand{\eea}{\end{eqnarray}}
\def\eq#1{eq.$\,$(\ref{#1})}

\newcommand{\AmS}{{\protect\the\textfont2
  A\kern-.1667em\lower.5ex\hbox{M}\kern-.125emS}}

\title{Reconstruction of the Proton Source in Relativistic Heavy 
       Ion Collisions}

\author{Alberto Polleri\address{The Niels Bohr Institute, Blegdamsvej 17, 
        DK-2100 Copenhagen \O, Denmark.}
        \thanks{Present address:  Institute for Theoretical Physics,
        Philosophenweg 19, 69120 Heidelberg, Germany.} ,
        Raffaele Mattiello$^a$ , 
        Igor N. Mishustin$^{a,}$\address{The Kurchatov Institute, Russian 
        Scientific Center, Moscow 123182, Russia.}$^{,}$\address{Institute for 
        Theoretical Physics, J.W.Goethe--University,
        60054 Frankfurt, Germany} and Jakob P. Bondorf$\,^a$.}

\begin{document}

\maketitle

\begin{abstract}
We describe a direct method to reconstruct the transverse proton source formed
in a relativistic heavy ion collision, making use of experimentally measured 
proton and deuteron spectra and assuming that deuterons are formed via 
two-nucleon coalescence. We show that an ambiguity with respect to the source 
temperature still persists and we indicate a possible solution to the problem. 
\end{abstract}

\vspace{4mm}

It has been recently shown that a simple description of the proton phase space
distribution can provide a good qualitative understanding of deuteron spectra
\cite{PBM98}.
On the other hand, it appeared clear that a more solid 
method to extract the properties of the source had to be established. We have 
therefore reconstructed the phase space distribution of protons directly from 
the observed proton and deuteron spectra, exploiting the coalescence 
prescription together with the notion of collective flow. Here we will outline
the procedure while all details can be found in \cite{PMMB99}.
We make use of a relativistic description of collective 
flow based on the boost-invariant picture for the longitudinal expansion, 
together with a longitudinally-independent transverse velocity, and
therefore write the proton phase space distribution as
\be
f_p(x,p) = (2\pi)^3 \exp(-\,p_\mu 
u^\mu(x)\,/\, T_0)\, B_p\, n_p(r_\perp)\,,
\label{PHSP}
\ee
where $p_\mu u^\mu(x) = \gamma(r)\, (m_\perp \cosh(y-\eta) - 
\vec{p}_\perp \cdot \vec{v}(\vec{r}))$ is the energy in the global frame and 
$B_p$ is the normalisation coefficient of the Boltzmann distribution in 
the local frame. The local density $n_p(r)$ is assumed to be independent of 
the longitudinal rapidity.

The deuteron phase space distribution is calculated using the
coalescence model. Its evaluation is simplified when considering large
and hot systems, neglecting the smearing effect of the 
deuteron Wigner density in comparison to the characteristic scales of the 
system in position and momentum space. One therefore obtains the
deuteron phase space distribution
\be
f_d(x,p) \simeq 
\mbox{\small $\frac{3}{8}$} R_{np} \left[ f_p(x,p/2) \right]^2\, .
\label{COAL}
\ee
The neutron to proton ratio in the source was taken to be $R_{np} = 1.2$.
The deuteron phase space distribution has the same structure as the
proton one in \eq{PHSP}, now with $n_p(r)$ replaced by $n_d(r) = 
\lambda_d\,n_p^2(r)$, where $\lambda_d = 3/8\,R_{np}(2\pi)^3\,B_p^2/B_d$.

One can now calculate proton and deuteron invariant momentum spectra
using the Cooper-Frye formula, on an approximate freeze-out 
hypersurface of constant 
longitudinal proper time $\tau_0$. The integrations over the space-time
rapidity and the azimuthal angle can be easily performed, yielding an 
expression in terms of the two functions $v(r)$ and $n(r)$. The ambiguity in 
the description of the single particle spectrum is explicit, since the two 
functions cannot be mapped out uniquely from only one function as the 
transverse momentum spectrum. To partially remove the ambiguity, we first
change the integration variable and introduce the auxiliary function 
$\tilde{n}$ through the relation $v\,dv\,\tilde{n}(v) = r\,dr\,\tau_0\,n(r)$, 
obtaining the new expression for the momentum spectrum
\be
S(p_\perp) = 4\pi\,B\,m_\perp \!\int_0^1\! dv v\, 
K_1\left(\frac{\gamma\, m_\perp}{T_0}\right)\,
I_0\left(\frac{v\, \gamma\, p_\perp}{T_0}\right)\,\tilde{n}(v)\,.
\label{NOAMBIG}
\ee
The one-to-one correspondence between $\tilde{n}(v)$ and $S(p_\perp)$ is now 
evident. We then make use of the coalescence model and
the definition of $\tilde{n}$, both for protons and for deuterons, to
obtain a first order differential equation that can be directly integrated 
and gives the closed solution
\be
r^2 = 2\, \frac{\lambda_d}{\tau_0} \int_0^{v}\!\! du\,u\, 
\frac{\tilde{n}_p^2(u)}{\tilde{n}_d(u)}\,.
\label{FLOWREC}
\ee
Therefore, by independently extracting the functions $\tilde{n}_p$ and 
$\tilde{n}_d$ from the observed momentum spectra $S_p(p_\perp)$ and 
$S_d(p_\perp)$, we can find the function $r(v)$ by a simple numerical 
integration. Inverting the obtained function as $r(v) \rightarrow v(r)$,
we obtain the collective velocity profile. We also obtain the local proton 
density as
\be
n_p(r) = \frac{v(r)}{r}\ \frac{dv(r)}{dr}\ \frac{\tilde{n}_p(v(r))}{\tau_0}\, .
\label{DENSREC}
\ee

\begin{figure}[t]
\begin{minipage}[t]{82mm}
\centerline{\psfig{figure=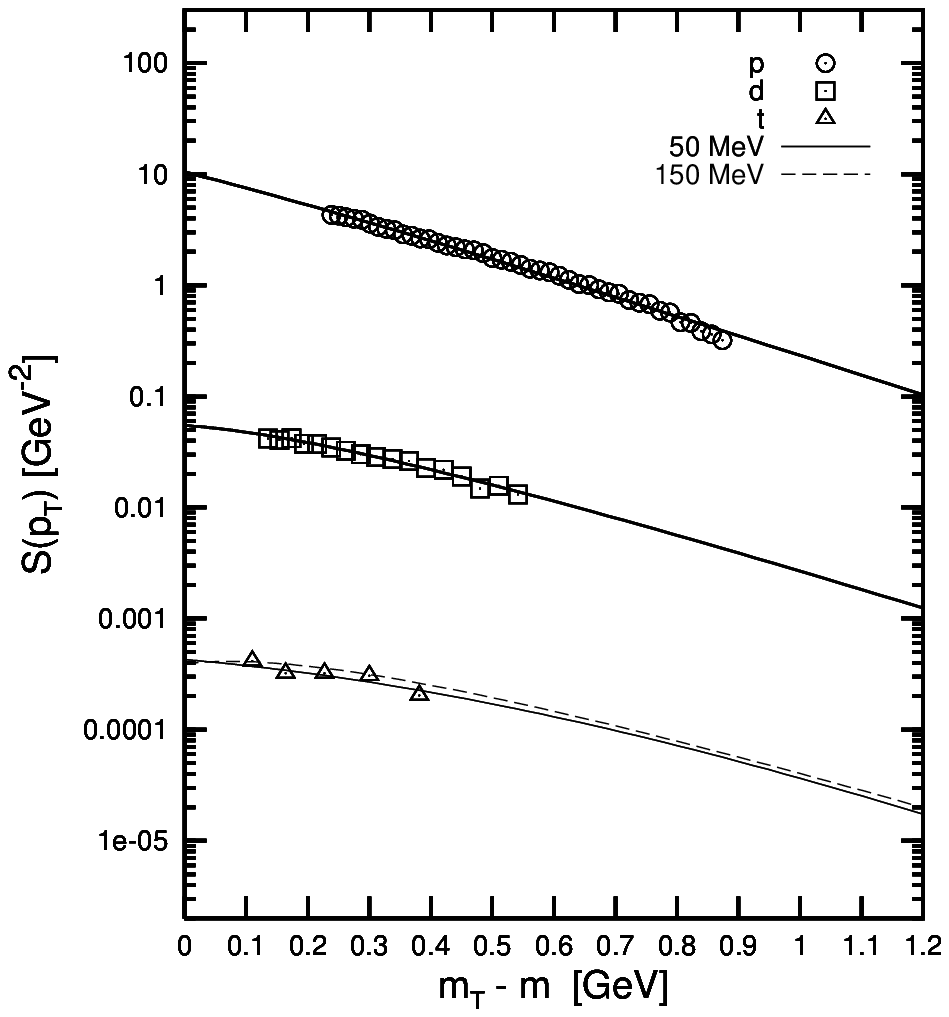,width=82mm,height=9cm}}
\protect\caption{Comparison of the transverse spectra of $p$, $d$ and $t$,
calculated within the improved coalescence model, with the measurements
obtained by the NA44 experiment \protect\cite{NA4499}.}
\label{SPECTRA}
\end{minipage}
\hspace{\fill}
\begin{minipage}[t]{74mm}
\centerline{\psfig{figure=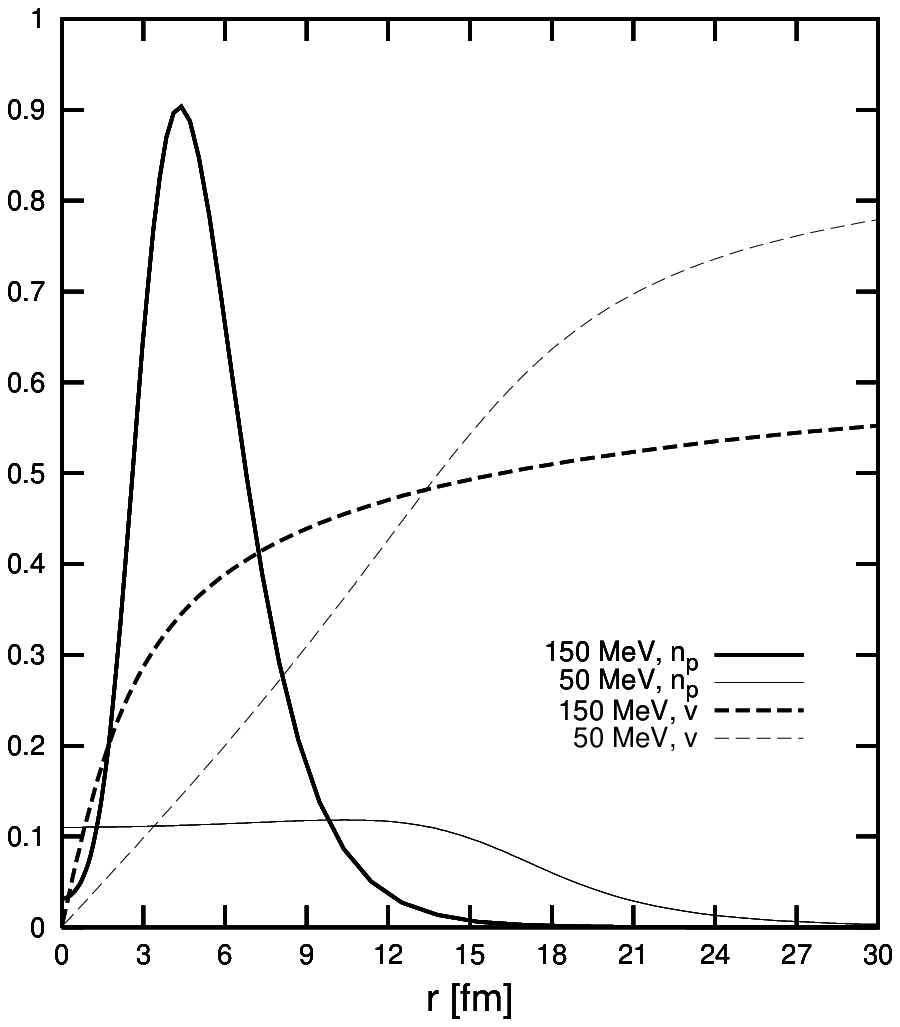,width=74mm,height=9cm}}
\protect\caption{Reconstructed local density and flow of protons as 
         function of the transverse radius, for the extreme values
         of $T_0$. $n_p$ is rescaled to be compared with $v$.}
\label{FUNCTIONS}
\end{minipage}
\end{figure}

The described procedure was applied to the transverse momentum spectra, 
resulting from Pb+Pb collisions and measured by the NA44 collaboration at the 
CERN-SPS \cite{NA4499}. We fitted $\tilde{n}_p$ and $\tilde{n}_d$ to these 
data, using eq.$\,$(\ref{NOAMBIG}) for protons and deuterons and assuming 
different values for $T_0$, from $50$ MeV to $150$ MeV. We chose a form of 
the profile functions 
characterized by three parameters, and we extracted their
value from the experimental data with a Monte Carlo search
minimising $\chi^2$. The fitted spectra are shown, for the extreme values of 
temperature considered, in the top part of 
Fig.$\,$\ref{SPECTRA}. Although evaluated at different temperatures, 
they are indistinguishable from one another. On the other hand, the profiles 
$\tilde{n}$ result to be very different for
different temperatures. 

In all calculations the freeze-out time was fixed at $\tau_0 = 10$ fm/c 
\cite{BBGH97}. After numerical integration of eq.$\,$(\ref{FLOWREC}), we 
obtained the function $v(r)$ shown in Fig.$\,$\ref{FUNCTIONS} for the two
extreme temperatures. It shows a linear rise at small $r$ and saturates
for large $r$. The velocity profiles clearly depend on the temperature chosen.
The local proton density is
also plotted in Fig.$\,$\ref{FUNCTIONS}. At high temperature the density shows
a shell-like structure which disappears as the temperature is lower. 
Similar shell-like structures have been recently found in analytic solutions 
of non-relativistic hydrodynamics \cite{C98}. From the plots
one can observe the different transverse sizes corresponding to different
temperatures. It is therefore necessary to know $T_0$ precisely
in order to determine the system size. This information cannot be  extracted 
solely from proton and deuteron spectra. Contrary to what is commonly done, 
source radii {\it cannot} be extracted from the $d/p^2$ ratio ($B_2 \sim $
emitting volume) unless $T_0$ is known.

It is clear that to resolve the remaining ambiguity in the source temperature
one needs some additional experimental information. Recently, $\pi\,\pi$ 
correlation data were used as a constraint \cite{SH99}. Since pions may 
freeze-out in a different way than protons, it would be even better to
consider $p\,p$ correlations, although they are more sensitive to 
final state interactions than pions. On the other hand, heavier clusters 
can provide additional constraints. We now address this issue following 
\cite{MSSG97} and we describe the fusion process of $A$-nucleons into
a bound state within the density matrix formalism. Making use of the same 
approximation leading to \eq{COAL},
one obtains cluster phase space distribution
\be
f_c(x,p) \simeq
g_A\,\frac{(R_{np})^N}{Z!\,N!}\,[f_p(x,p/A)]^A\,.
\label{FULLCOALMOD}
\ee
For $N=Z=1$, $A=2$ and $g_2=3/8$ it reduces to \eq{COAL}.
The statistical prefactor $g_A$ is of crucial importance. In the conventional 
approach we have 
$g_A = (2S_A + 1)Z!\,N!\ /\ 2^A A!$, which results in 
$g_d = 3/8$, $g_t=g_{^3\!He}=1/12$,
and $g_{^4\!He}= 1/96$.
Although straightforward, this approach has been successful only for the 
description of deuteron production. In fact, it significantly underestimates
the yields of heavier clusters. A possible improvement can be achieved by 
allowing for additional formation processes. Besides the already considered 
direct process, it is possible that deuteron-like correlations contribute to 
cluster formation. For the triton, as an example, we should therefore account 
for the possibility that a proton and a neutron are already in a bound state 
with deuteron quantum numbers and coalesce with another
neutron, with a different statistical prefactor. More precisely we can write
$g_t = g_{pnn \rightarrow t} + 2 g_{pn \rightarrow d}\, g_{dn \rightarrow t}$,
where the factor 2 counts the different ways to associate the proton with the
two neutrons in forming a deuteron. The spin-isospin counting is 
straightforward and gives the modified statistical prefactor 
$g_t = 1/3$, therefore enhancing the triton yield by a factor 4. The
same arguments apply to $^3\!He$, so that $g_{^3\!He} = 1/3$. The case of
$^4\!He$ is more involved. Counting all the possible processes we obtained
$g_{^4\!He} = 13/48$,
so that the $^4\!He$ yield is increased by a factor 26. For more details 
see \cite{ALBPHD99}.

We now use the coalescence model in this improved version to calculate the 
transverse mass spectrum of tritons. Using the flow and 
density profiles extracted from the analysis of $p$ and $d$ spectra, we 
examine to what extent heavier clusters can constrain the 
ambiguity of the temperature. Using eqs.$\,$(\ref{PHSP}) and 
(\ref{FULLCOALMOD}), we obtain $n_t(r) = \lambda_t \, n_p^3(r)$, with
$\lambda_t = g_t (R_{np})^2/2\, (2\pi)^6\, B_p^3/B_t$, while
the collective velocity is the same for all clusters. The triton spectrum is
plotted in Fig.$\,$ \ref{SPECTRA}, together with the previously fitted $p$ and
$d$ spectra. The absolute values and the shape compare very well with the 
experimental data. This confirms, that the improved statistical approach is 
consistent with the measured spectra. Furthermore, the results obtained
with the two extreme temperatures show a different curvature.
The high temperature case presents a clear bending over, absent for low 
temperature. We argue that this specific difference might narrow down the 
allowed temperatures and therefore provide additional constraints.
Unfortunately, the triton spectrum was measured only in a limited range in
transverse momentum, and therefore a quantitative fit is not useful.

We also suggest that a further possibility to constrain the temperature lies
in the combined study of single and composite spectra, together with $p\,p$ 
correlations. The extracted profiles shown in Fig.$\,$ \ref{FUNCTIONS} could 
in fact be used to evaluate the $p\,p$ correlation function, which is sensitive
to temperature and flow in a different way respect to inverse slopes of 
spectra. These remarks are important for the interpretation of the future 
experiments at the Relativistic Heavy Ion Collider.

\end{document}